\documentclass[twocolumn]{aa}
\usepackage{graphicx}
\usepackage{txfonts}
\begin{document}

\newcommand {\flux} {{$\times$ 10$^{-11}$ erg cm$^{-2}$ s$^{-1}$}}
\def\magcir{\raise -2.truept\hbox{\rlap{\hbox{$\sim$}}\raise5.truept
\hbox{$>$}\ }}

   \title{BeppoSAX observations of Seyfert 1s in the Piccinotti sample I: poorly studied sources}
%   \subtitle{ }

   \author{A.~Quadrelli\inst{1}, A.~Malizia \inst{1}, L.~Bassani\inst{1}, G.~Malaguti\inst{1}}

   \offprints{A. Quadrelli (quadrelli@bo.iasf.cnr.it)}

   \institute{IASF/CNR, via Piero Gobetti 101, I-40129 Bologna, Italy}
   \date{Received / accepted}

   \titlerunning{BeppoSAX observations of Seyfert 1s in the Picinotti sample I: poorly studied sources}

   \authorrunning{A. Quadrelli et al.}

\abstract{In this work we present the first of two papers devoted to
the study of the X-ray spectral characteristics of Seyfert 1 galaxies
in the Piccinotti sample. In particular we analyse here the BeppoSAX
broad band (0.1-100 keV) data of 4 objects which, despite their X-ray
brightness, have been historically poorly studied due to their late
identification with an AGN; these are H0111-149 (MKN1152), H0235-525
(ESO198-G24), H0557-385 (IRAS F05563-3820) and H1846-786 (IRAS
F18389-7834).  We have assumed for all the sources a baseline model
which includes a power law with an exponential cut-off plus a
reflection component and an iron K$_\alpha$ line; we have also
searched for the presence of intrinsic absorption and/or a soft excess
component.  Our analysis indicates the presence of complex absorption
in two objects (H0557-385 and H0111-149) best described by a
combination of two uniform absorbers, one cold and one warm. Only in
one source, H0557-385, a soft excess component has been measured. The
primary continuum is best described by a canonical power law
($\Gamma$=1.7-2) with a high energy cut-off in the range 40-130 keV.
A cold reflection component is likely present in all sources with
values ranging from less than 0.6 to higher than 2. In 3 out of 4
objects we find a cold iron line having equivalent width typical of
Seyfert 1s (100-200 eV).

}

 \maketitle

%________________________________________________________________

\section{Introduction}
The A2 experiment on the HEAO-1 satellite performed a complete X-ray
survey of 8.2 sr of the sky (65.5\% coverage) at $|b|\ge20^{\circ}$ in
the 2-10 keV band (Piccinotti et al. 1982).  Among the 85 sources
detected down to a limiting flux
of $3.1\times10^{-11}$ erg cm$^{-2}$ s$^{-1}$, 36 were classified as
active galactic nuclei (AGN). They form a complete hard X-ray selected sample
of nearby AGN consisting of 22 Seyfert galaxies of type 1-1.5, 8 of
type 1.8-2 plus 1 QSO (3C273), 1 Starburst galaxy (M82) and 4 BL Lac
objects.  The sample of type 1 Seyferts is big enough to allow some
statistical studies of the bulk properties of these type of AGN: by
analysing the X-ray spectra of the entire sample we can tackle a
number of issues still unsettled, like defining the incidence of soft
excesses, the role and type (cold and/or warm) of absorption(s), the parameter
space of various spectral components such as the reflection, the high energy
cut-off, the photon index and the iron line equivalent width (EW) as well
as the presence of correlations between them.

BeppoSAX observations offer in this sense a unique opportunity since they
provide broad band data of sufficiently high quality to allow such a
study.  Of the 21 objects in the sample, BeppoSAX has observed 18 sources over
the entire range of the instrument (0.1-200 keV), while one (NGC3227)
has been serendipitously detected by the PDS instrument only.  

Here we present the first of two papers devoted to the study of the broad band
X-ray spectral characteristics of Seyfert 1 galaxies in the Piccinotti
sample: in particular, we analyse 4 objects which, 
despite their X-ray brightness,  have been historically poorly studied  due to their late
identification with an AGN.  These are H0111-149 identified with
MKN1152 at z=0.0536 by Turner and Pounds (1989), H0235-525 identified
with ESO198-G24 at z=0.045 by Ward and Shafer (1988), H0557-385
identified with IRAS F05563-3820 at z=0.034 by Fairall, McHardy $\&$
Pye (1982) and H1846-786 identified with IRAS F18389-7834 at z=0.084
by Remillard et al. (1986). 
X-ray spectra of these objects were obtained for the first time by
EXOSAT (Turner and Pounds 1989), except for H0557-385, since in this
case the data were contaminated by the presence of a nearby source
(Giommi et al 1989). The observed spectra were all canonical power
laws with evidence of intrinsic absorption in one (H0111-149) out of 3
objects.  All 4 sources were observed by ASCA but only in the case of
H0557-385 the data have been published (Turner et al. 1996).  

\begin{table*}
\begin{center}
\centerline{{\bf Table 1: Observation Log}}
\begin{tabular}{lcccccccc}
\hline\hline 
{\it Source} & {\it Start date}  & \multicolumn{3}{c} {\it $T_{exp}$ (Ks)} & &  \multicolumn{3}{c}{\it Counts/s }\\
\cline{3-5} \cline{7-9}
             &                   & {\it LECS}        & {\it MECS}   & {\it PDS}  &           & {\it LECS} & {\it MECS} & {\it PDS}\\
\hline
H0235-525 (1)   & 2001-Jan-23  & 56   & 144              & 63        &       & 0.13$\pm$0.002  & 0.20$\pm$0.001 & 0.32$\pm$0.02 \\
H0235-525 (2)   & 2001-Jul-05  & 23   & 97               & 44        &       & 0.07$\pm$0.002  & 0.11$\pm$0.001 & 0.23$\pm$0.03 \\
H0557-385 (1)   & 2000-Dec-19  & 21   & 29               & 15        &       & 0.23$\pm$0.004  & 0.44$\pm$0.004 & 0.47$\pm$0.05  \\
H0557-385 (2)   & 2001-Jan-26  & 4.8  & 14               & 5.8       &       & 0.29$\pm$0.009  & 0.51$\pm$0.006 & 0.51$\pm$0.06  \\
H0111-149       & 2001-Jan-04  & 33   & 79               & 40        &       & 0.05$\pm$0.001  & 0.07$\pm$0.001 & 0.11$\pm$0.03  \\
H1846-786       & 2001-Mar-08  & 85   & 34               & 18        &       & 0.08$\pm$0.004  & 0.11$\pm$0.002 & 0.21$\pm$0.04 \\
\hline\hline
\end{tabular}\end{center}
\end{table*}

\begin{table*}
\begin{center}
\centerline{{\bf Table 2: Fluxes}}
\begin{tabular}{lccccc}
\hline\hline
{\it Source}        &  \multicolumn{3}{c} {\it $Flux^{(a)}$}                            &  {\it $N^{(b)}_{Gal}$}   & $z$                \\
\cline{2-4}
                    &  {\it (0.1-2 keV)}   & {\it (2-10 keV)}    & {\it (20-100 keV)}   &            &             \\
\hline										      
H0235-525 (1)     &  0.37                &  1.5                &  2.4                   &  3.05      & 0.0455      \\
H0235-525 (2)     &  0.07                &  0.9                &  2.2                   &  3.05      & 0.0455     \\
H0557-385 (1)     &  0.8                 &  3.6                &  2.8                   &  3.98      & 0.0344      \\
H0557-385 (2)     &  1.0                 &  4.0                &  5.7                   &  3.98      & 0.0344    \\ 
H0111-149         &  0.11                &  0.57               &  0.99                  &  1.60      & 0.0527     \\ 
H1846-786         &  0.38                &  0.76               & -                      &  9.06      & 0.0740     \\
\hline\hline
\end{tabular}
\end{center}
$^a$ in units of $10^{-11}$erg $cm^{-2}$ $s^{-1}$.\\
$^b$ Galactic column density in units of 10$^{20}$ cm$^{-2}$
\end{table*}

On top of a canonical AGN spectrum, H0557-38 presented a complex structure
below 2 keV indicative of attenuation by an ionized absorber fully or
partially covering the source, beyond a neutral absorber fully
covering the source; a soft excess component was also evident. Finally
an iron K-shell emission line, which appeared to be significantly
broad, with an equivalent width of 300 eV was detected.  Inspection of
the Tartarus database (see http://tartarus.gsfc.nasa.gov) for
unpublished observations, confirms EXOSAT findings of a canonical
power law in the other 3 objects, the presence of intrinsic absorption
in H0111-149 and evidence for an iron line in H0111-149 and H0235-525.
So far BeppoSAX data have been published for only one source, i.e. the
first of the two pointings performed on H0235-525. Guainazzi (2003)
discussed these data in conjunction with an XMM measurement of the
source, which had a spectrum characterized by reprocessing features
(reflection and iron line) produced in an accretion disk. 

At low energies (0.1-2 keV ) the entire sample has been studied by Schartel
et al. (1997) using all sky survey data from the ROSAT
satellite. These authors found that all 4 objects could be fitted with
power laws having photon indeces in the range 2.1-3.2 and that only
one source (H0557-385) displayed significant absorption.  

Finally, Malizia et al. (1999) performing a systematic coverage of the whole
Piccinotti sample with BATSE have indicated that all four sources were
marginally detected (about 3$\sigma$ level) at high energies with a
20-100 keV flux in the range 4-6 $\times10^{-11}$ erg cm$^{-2}$
s$^{-1}$.  Long term light curves in the 2-10 keV band indicate
substantial flux variations in all 4 sources over timescales of
months/years and so BeppoSAX very likely sampled a particular state of
each source.

\section{Data analysis}
The BeppoSAX observation Log is shown in Table 1 which lists for each
source the observation date, the exposure time and the count rate in each
of the narrow field instruments, while other relevant data such as the source
redshift, the galactic column density in the direction of each object and the
0.1-2, 2-10 and 20-100 keV fluxes relative to the best fit model are
listed in Table 2. 
 
Standard data reduction was performed using the
software package "SAXDAS" (see http://www.asdc.asi.it/software and the
Cookbook for BeppoSAX NFI spectral analysis, Fiore, Guainazzi \&
Grandi 1998).  Data were linearized and cleaned from Earth occultation
periods and unwanted period of high particle background (satellite
passages through the South Atlantic Anomaly).  Data have been
accumulated for Earth elevation angles $>5$ degrees and magnetic
cut-off rigidity $>6$. For the PDS data we adopted a fine energy and
temperature dependent Rise Time selection, which decreases the PDS
background by $\sim 40 \%$. This improves the signal to noise ratio of
faint sources by about 1.5 (Frontera et al. 1997, Fiore, Guainazzi \&
Grandi 1998).

\begin{figure*}
\centering
\hbox{
\includegraphics[width=4.0cm,height=9.0cm,angle=-90]{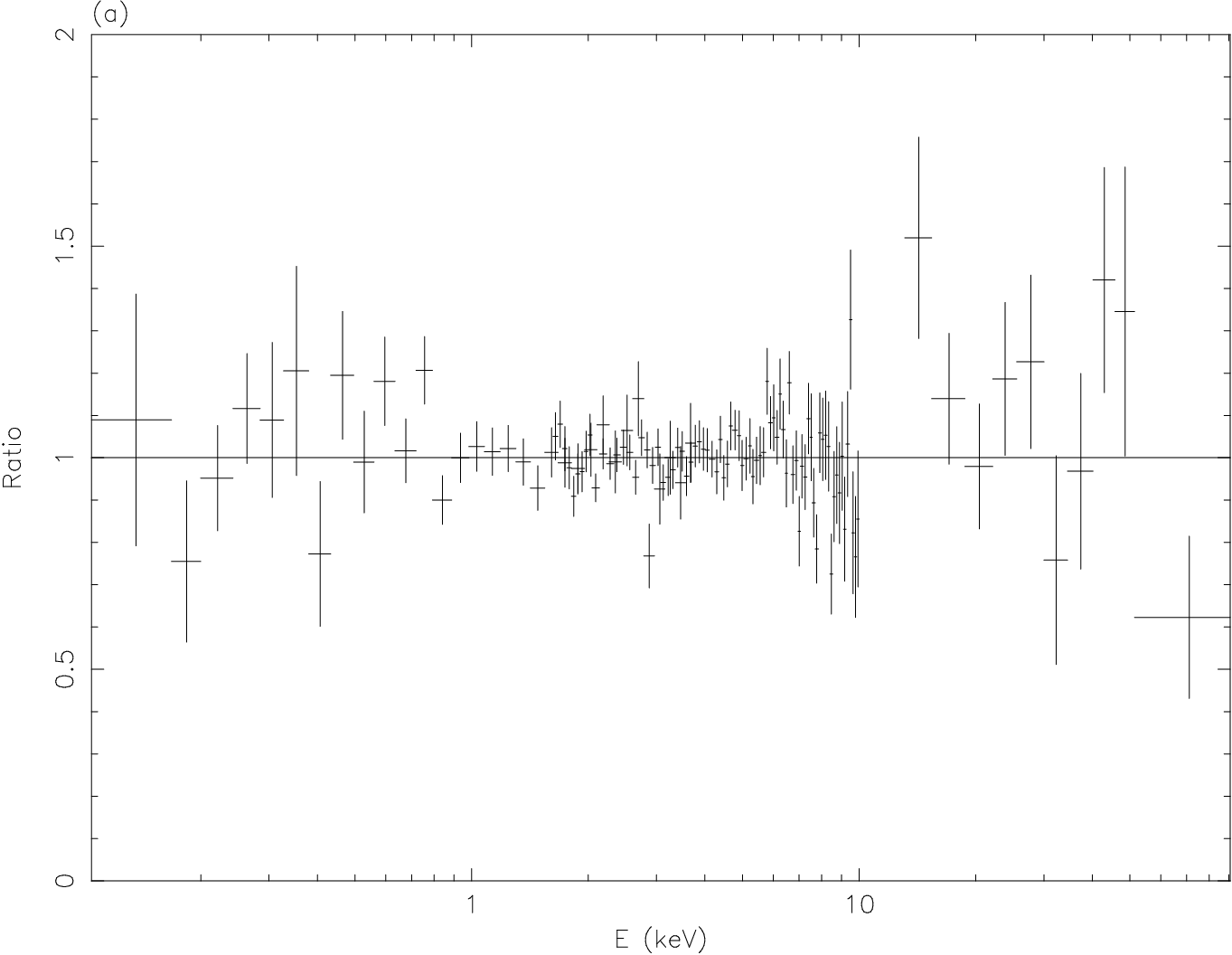}
\includegraphics[width=4.0cm,height=9.0cm,angle=-90]{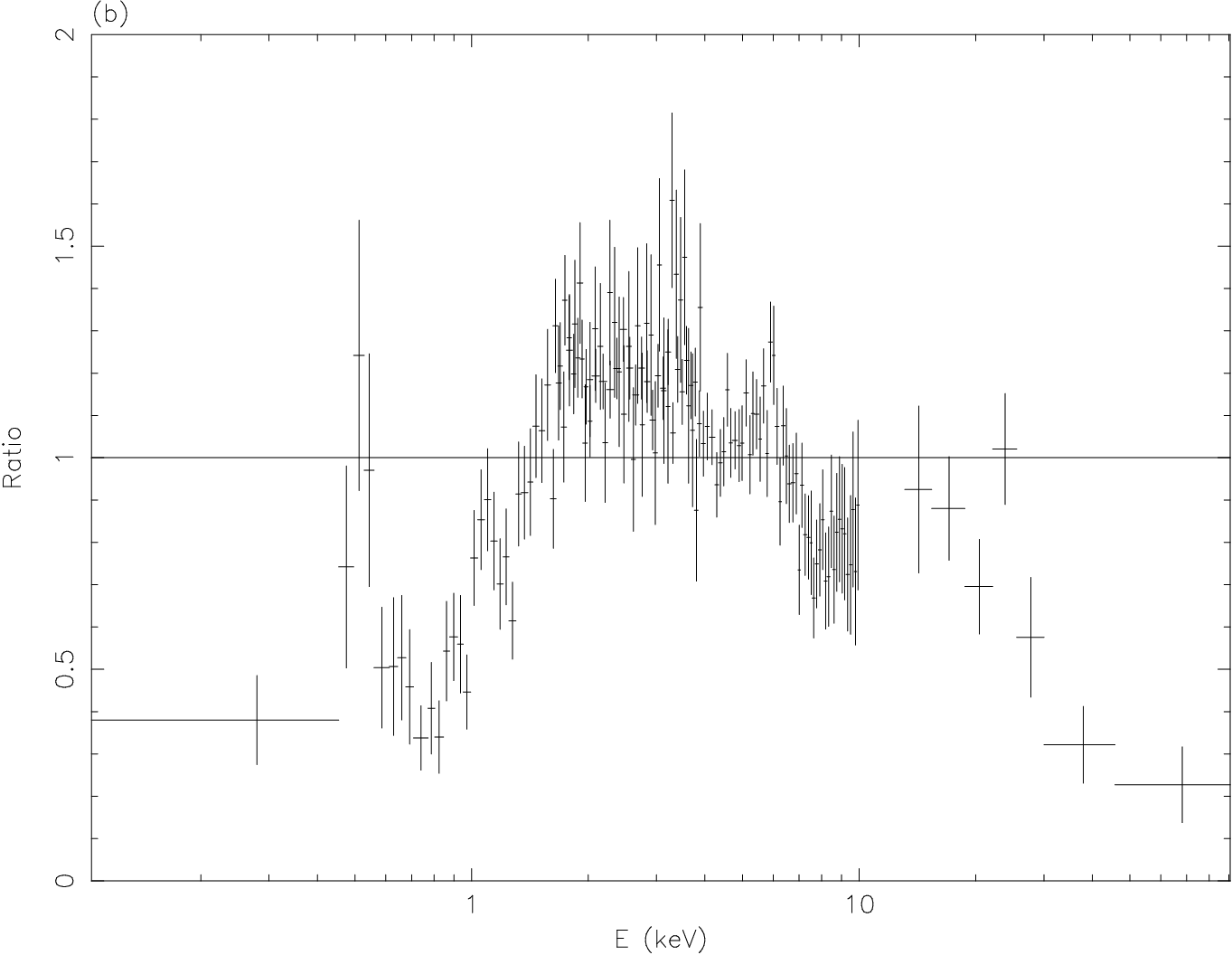}
}
\end{figure*}

\begin{figure*}
\centering
\hbox{
\includegraphics[width=4.0cm,height=9.0cm,angle=-90]{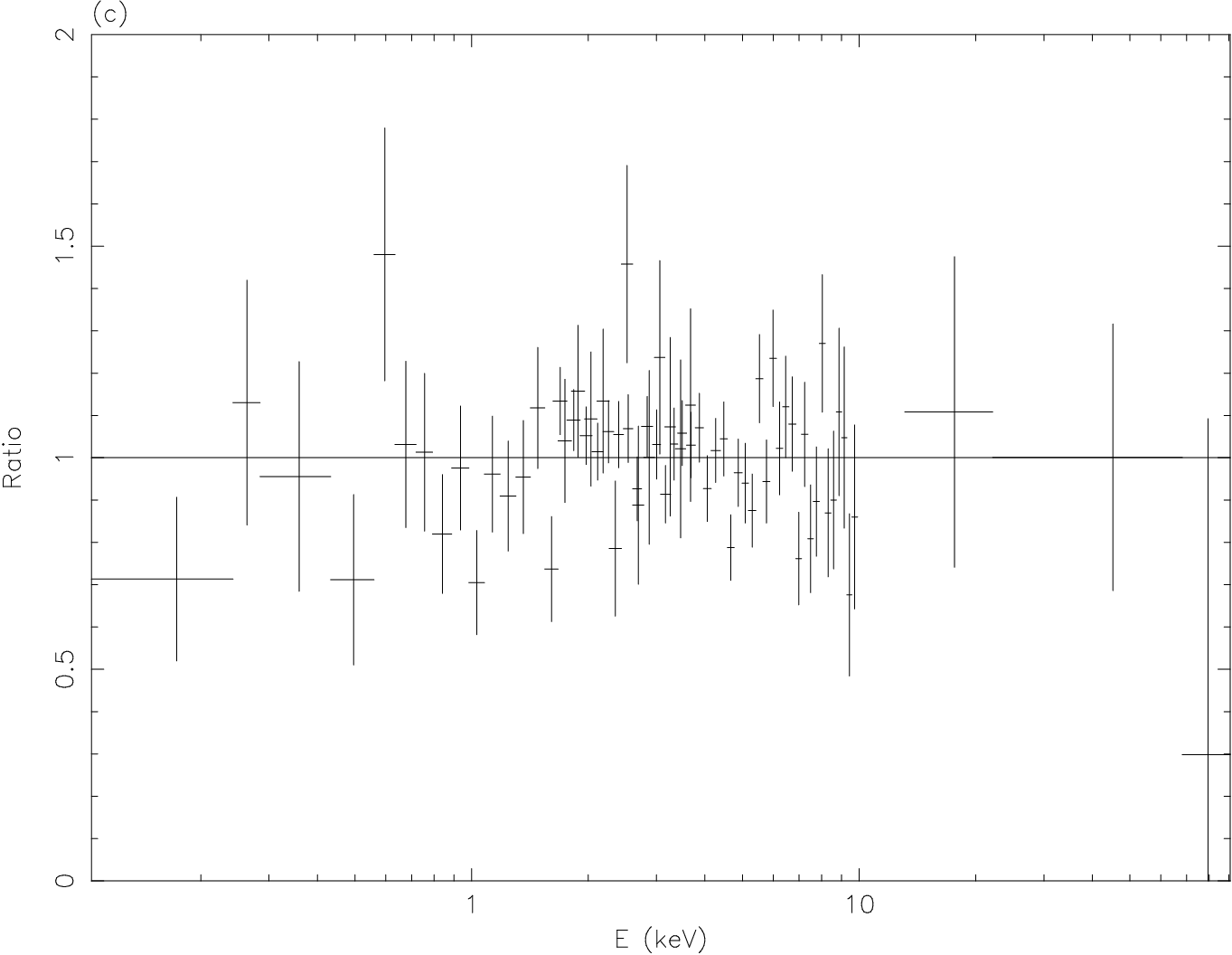}
\includegraphics[width=4.0cm,height=9.0cm,angle=-90]{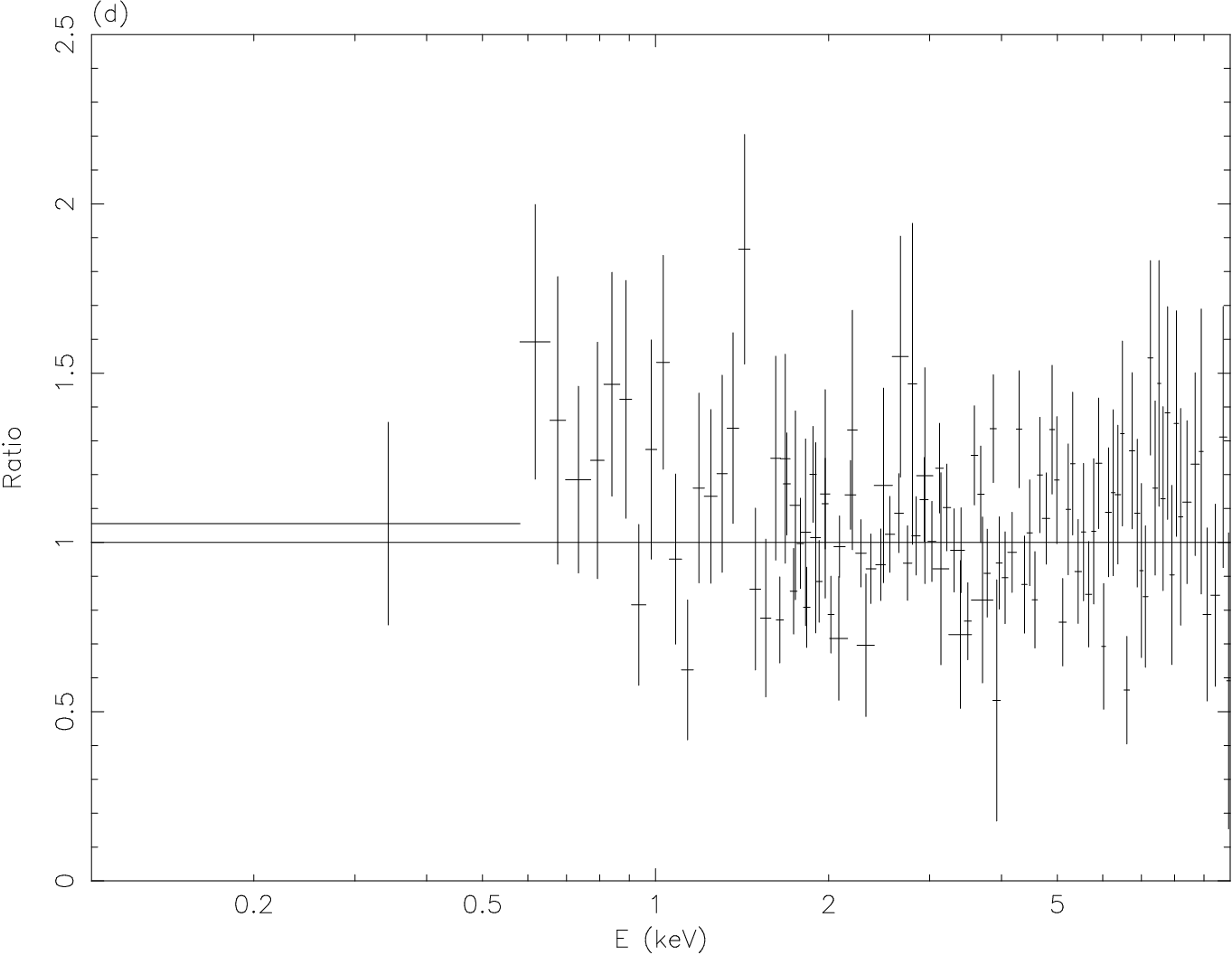}
}
\caption[]{Data to model ratios of H0235-525 during the second
observation (a), H0557-385 during the first observation (b), H0111-149
(c) and H1846-786 (d).The fitting model is a simple power law absorbed
by the Galactic column density.}
\end{figure*}

\subsection{Imaging Analysis}
Inspection of the LECS and MECS images indicates the
presence of serendipitous X-ray sources in the field of view of two of
our sources: H0111-149 and H0557-385.  In the first case, an active
galaxy identified in the Simbad catalog as EXO 0111.4-1458 (1RPX
J01135601442.4) is visible in both LECS and MECS images and located at
only 8.1 arcmin from H0111-149. Its count rate in the 0.1-2 keV is
20 times lower than that of H0111-149, thus excluding contamination in
the PDS energy band.  In the case of H0557-385 two sources are
detected in the LECS image, one of which is also visible in the
MECS field of view. The object detected by both instruments is identified with a
cataclysmic variable star 30 arcmin away from the target source; the
object only seen by the LECS is instead a BL Lac object
(EXO055625-3838.6) 25 arcmin from H0557-385.  Both sources have an
X-ray spectrum too steep (the star is not detected above 5 keV and the
BL Lac above 2 keV) to be considered as potentially contaminating
for the PDS. No other objects, beside the target source is
visible in the images of H0235-525 and H1846-786.
Despite this, the MECS-PDS cross calibration constant obtained in
the fit of H1846-786 suggests the
presence of one or more contaminating sources in the PDS observation
of this Seyfert; this constant is  much higher (2.03-3.85)
than the expected value of 0.70-0.95. 
Within 1 degree of H1846-786, we find
a bright Seyfert 1 galaxy (ESO025-G002, z=0.028), having a ROSAT flux only 5
times lower than the target object; at least 3 more AGN of type 1 are present in the
PDS field of view although they are all much less intense. 
For this reason, special caution
was taken when treating the PDS data of H1846-786 (see section 2.4).

\begin{table*}
\begin{center}
\centerline{{\bf Table 3: Results of the spectral analysis.}}
\begin{tabular}{lccccccc}
\hline\hline 
{\it Source}  & {\it $\Gamma$}  & {\it $E_c$}  & {\it R}            & {\it E (Fe$_{K\alpha}$)}  & {\it EW}    & {\it cos i} & {\it $\chi^2$/d.o.f.}\\
              &                 & {\it (keV)}  & {\it (cos i = 0.9)}  & {\it (keV)}         & {\it (eV)}  & {\it (R=1)} & \\
\hline
H0235-525 (1)   & $1.74^{+0.03}_{-0.03}$ & $132^{+441}_{-64}$ & $0.49^{+0.41}_{-0.34}$ & $6.35^{+0.34}_{-0.22}$ & $61^{+35}_{-36}$ & $0.25^{+0.16}_{-0.11}$ & 128 / 117 \\
H0235-525 (2)   & $1.72^{+0.02}_{-0.04}$ & $128^{+114}_{-45}$ & $1.50^{+0.39}_{-0.32}$ & $6.68^{+0.31}_{-0.32}$ & $120^{+67}_{-69}$ & $0.90^{+0.05}_{-0.25}$ & 118 / 110 \\
H0557-385 (1)   & $1.92^{+0.04}_{-0.03}$ & $41^{+80}_{-9}$    & $1.66^{+0.47}_{-0.36}$ & $6.17^{+0.13}_{-0.14}$ & $121^{+52}_{-52}$ & $0.90^{+0.05}_{-0.13}$ & 142 / 165 \\
H0557-385 (2)   & $2.02^{+0.04}_{-0.04}$ & $>130$             & $1.34^{+0.36}_{-0.41}$ & $6.49^{+0.40}_{-0.34}$ & $<152$ & $0.89^{+0.06}_{-0.46}$ & 85 / 73 \\
H0111-149       & $1.71^{+0.15}_{-0.10}$ & -                  & $<0.57$                & $6.49^{+0.54}_{-0.34}$ & $130^{+95}_{-98}$ & $<$0.33 & 79 / 76 \\
H1846-786       & $1.92^{+0.36}_{-0.04}$ & -                  & -                      & 6.4 (fixed)            & $<205$ & - & 160 / 142 \\ 
\hline\hline
\end{tabular}\end{center}
\end{table*}

\subsection{Spectral Analysis}
Spectral data were extracted from regions centered on each source with
a radius of 4 arcmin for LECS and MECS; background spectra were taken
from blank sky field regions having the same size in detector
coordinates.  Source plus background light curves did not show any
significant variation and therefore data have been grouped for
spectral analysis.  Spectral fits were performed using  XSPEC
11.2.0 software package and public response matrices as from the 1998
November issue.  PI (Pulse Invariant) channels were rebinned
in order to have at least 20 counts per bin.  This guarantees the use of the $\chi^2$ method in
determining the best fit parameters, since the distribution in each
channel can be considered Gaussian.  Constant factors have been
introduced in the fitting models in order to take into account the
inter-calibration systematic uncertainties between instruments (Fiore,
Guainazzi \& Grandi 1998).  All  quoted errors correspond to 90\%
confidence level for one interesting parameter ($\Delta\chi^2$ =
2.71).  All  models used in what follows contain an additional term
to allow for the absorption of X-rays due to our galaxy (see column
5 of Table 2).
The $N_H$ galactic values, based on 21cm radio measurements, were provided by XSPEC.
 
The data to model ratios assuming a power law absorbed by only a galactic column density are shown in 
figure 1 which highlights spectral differences between sources and extra features to
be investigated further. To this end and also to compare our data to
previous works, we have assumed for all sources a baseline model which
includes: a photon power law (PL) with exponential cut-off
[A$E^{-\Gamma}$ exp(-E/$E_f$)] together with a reflection component (R=$\Omega/2\pi$)
from a cold slab isotropically illuminated by the PL photons and
subtending a solid angle $\Omega$, with inclination cos {\it
i} to the line of sight (Magdziarz \& Zdziarski 1995, module PEXRAV)
and a Gaussian line in the rest frame of the source to represent the
iron K$_\alpha$ fluorescence. In no case we found evidence for a broad line and
therefore in the following analysis the line was always assumed to be
narrow (sigma=0).  In addition to the Galactic column density we have
also considered intrinsic absorption at the source following the
results of Perola et al. (2002).  Furthermore, for each sample source
we have investigated the nature of this intrinsic absorption assuming
either a uniform cold absorber, a partial covering absorber
(model PCFABS) and/or a uniform warm absorber (model ABSORI).
For both reflection and absorption we adopted the element
abundances as in Anders $\&$ Grevesse (1993).  
All fits were performed
with cos {\it i} fixed  to 0.9 ({\it i} = $30^{\circ}$), i.e. appropriate for 
Seyfert of type 1.  
Alternatively  cos {\it i} was left free to vary, while R was
fixed equal to 1. Because the dependence on cos {\it i} of 
the reflection component shape is modest, all other parameters turned out identical 
in practice to those obtained with the fixed angle and the differences in $\chi^{2}$ 
are insignificant. 

Finally we investigated the presence of a soft excess in all  sources.
Some of the best fit parameters relative to adopted baseline model are listed in Table 3
together with the significance of the fit. Absorption and soft excess components when present are
discussed in the text. In the following sections results for each individual source are 
presented and discussed.

\subsection{H0235-525 (ESO198-G24)}
BeppoSAX-NFIs pointed at H0235-525 twice (January and July) during
2001 and found the source in two different states; consequently the
two measurements have been analyzed individually.  Fitting each
observation with our baseline model, we find compatible parameters for
the primary continuum: photon index $\Gamma\sim$1.7 and cut-off energy
E$_c$ $\sim$ 130 keV (see Table 3).  No evidence for intrinsic absorption and/or a
soft excess component is present in both data sets (see ratio of figure 1 (a)).  We measure an
iron line with high significance ($>$99$\%$ confidence level) but at
slightly different energies, $\sim$6.4 keV in the first observation versus
$\sim$6.7 keV in the second one. However, when we plot the iron
line energy versus normalization for the two measurements, we find that
both parameters are compatible within their relative uncertainties.
The only parameter which changes with some confidence
between the two BeppoSAX observations is the reflection
component. This is clearly seen in figure 2 where we plot R versus the
high energy cut-off: a factor of 3 higher reflection component is
present in the second observation when the 2-10 keV flux is also a
factor 1.7 lower than in the first measurement (see also tables 2-3).

The results obtained for the first observation are fully compatible 
with the analysis performed by Guainazzi (2003)  although we find a 
slightly lower equivalent width for the iron line (60 eV compared to 100 eV).
This author also studied the time history of the iron line in
H0235-525 analysing ASCA, BeppoSAX and XMM observations. He found that
during the XMM observation (December 2000) the line was broad and
twice as bright as in the BeppoSAX measurement (January 2001); by
contrast in the earlier ASCA observation (July 1997) the line was
dominated by a remarkably narrow core. This finding was interpreted
by the author as  indicative of  a time lag in the response of the 
line to the continuum flux in the same way  observed by us for the reflection component:
in other words, features due to  reprocessing emission are delayed in their 
response to changes in the source intensity. 
Considering that the time span between our two observations is of the
order of 160 days, we can infer an upper limit of 0.14 pc to the
distance between the primary continuum emission region and that where
reflection occurs.  This is compatible with the reprocessing region
being localized in the most external part of an accretion disk, in
the broad line region or even in a molecular torus. 

Combining our results with those of Guainazzi (2003) suggests that the last two
sites are more likely to be the region where reprocessing occur.

\begin{figure}
\centering
\includegraphics[width=7.0cm,height=9.0cm,angle=-90]{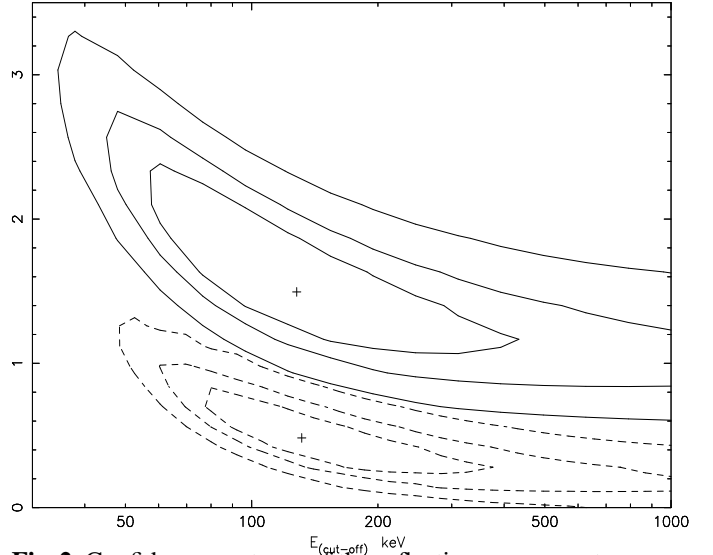}
\caption{Confidence contours of the reflection component versus the high energy cut-off in  H0235-525:
dashed lines are referred to the first observation and solid lines to the second.}
\end{figure}

\subsection{H0557-385 (IRAS F05563-3820)}
Also H0557-385 has been observed twice by BeppoSAX (December 2000 and
January 2001) and found at a similar flux level below 10 keV; above
this energy the flux changed by about a factor of 2 in just over a
month.  Therefore the two observations were  analyzed
individually to search for possible spectral differences particularly
at high energies.  It is evident that the simple power law model is
significantly inadequate to fit the data (see figure 1 (b)). But
this is hardly surprising giving that the spectrum below 2 keV is
highly complex as already shown by ASCA (Turner Netzer $\&$ George
1996) .  This source requires both intrinsic absorption and a soft
excess component. The absorber is complex with a cold and a warm
column density, both fully covering the source.  
The cold absorber is of the order
of 1.1$\pm$0.3 $\times$ 10$^{21}$ cm$^{-2}$ in both observations while the warm
absorber is N$_{H^*}=8.6^{+1.2}_{-0.9}$ $\times$ 10$^{21}$ cm$^{-2}$ with
$\xi$=5.9$^{+7.8}_{-4.8}$ erg cm s$^{-1}$ in the first observation and
N$_{H^*}=6.5^{+1.1}_{-0.9}$ $\times$10$^{21}$ $cm^{-2}$ with
$\xi$=4.3$^{+4.5}_{-3.2}$ erg cm $s^{-1}$ in the second one.
Despite the introduction of this complex absorption component, the fit
with our baseline model still leaves some residuals below 1 keV,
strongly indicating the presence of a soft excess component. Following
previous ASCA results, we assumed that this soft component maybe due
to emission from a hot plasma and parameterize it using a standard
Raymond-Smith model. The temperature of the plasma is found to be
0.25$\pm$0.2 keV, in agreement with what already observed with ASCA  by
Turner, Netzer $\&$ George (1996).
  
The primary power law is canonical ($\Gamma$=1.9-2)
in both observations with a high energy cut-off increasing from the
first (E$_{c_{1}}$=41 keV) to the second pointing (E$_{c_{2}}>$130 keV). 
Inspection of the contours of the cut-off energy versus the power 
law photon index for the two
observations, indicates that the values of E$_c$ are barely compatible 
at the 90$\%$ confidence level.
This change in the high energy cut-off energy also explains the
difference in flux above 10 keV.  
An iron line is detected in each measurement
with high significance ($\sim$ 99\% confidence level): the line is
compatible with being neutral and has an equivalent width value
typical of Seyfert 1 galaxies (EW= 120-150 eV).  
The reflection is quite strong (R$>$1) with values compatible between the two observations.

\subsection{H0111-149 (MKN 1152)} 
Although marginally evident in the ratio reported in figure 1(c),
H0111-149  requires absorption in excess to the galactic value at the
90$\%$ confidence level: an ionized absorber ($N_{H}=1.61^{+0.55}_{-0.47}$ $\times$ 10$^{21}$
cm$^{-2}$, $\xi$ 1.12$^{+1.68}_{-0.68}$ erg cm s$^{-1}$ ) or a cold one partially
covering the source (N$_H$=4.2$^{+1.0}_{-1.0}$ $\times$ 10$^{21}$ $cm^{-2}$, C$_{f}=0.40^{+0.30}_{-0.16}$\%) fit
equally well the data ($\chi^2_{warm}$=81 for 77 d.o.f  versus $\chi^2_{cold}$=82 for 77 d.o.f.).  
If instead we introduce
in the fits both a cold and warm absorber fully covering the source as
done in H0557-525 and used by Perola et al. (2002), the fit is
slightly better ($\chi^2$=79.3 for 76 d.o.f.) and the values for the cold absorber is 
$N_{H}=1.19^{+0.81}_{-0.59}$ $\times$ 10$^{20}$ $cm^{-2}$ while the warm one has 
$N_{H}=2.66^{+1.17}_{-2.49}$ $\times$ 10$^{21}$ at 
$cm^{-2}$ and  $\xi=20^{+34}_{-14}$
erg cm s$^{-1}$.  In Table 3 we report the best fit
parameters relative to this last case. 
No soft excess is required by the broad band fit.  
The primary power law is canonical ($\Gamma$=$1.71^{+0.15}_{-0.10}$) and  a neutral
iron line at $E_{line}$=6.5 keV is detected with high confidence (90\%) with an EW=125 eV .

Unfortunately, the high energy data are of too low statistical quality to allow good
constrain to be put on the reflection and high energy cut-off; fixing
the high energy cut-off at E$_c>$ 100 keV, we were able to constrain the
reflection to be smaller than 0.6.

\subsection{H1846-786 (IRAS F18389-7834)}
As discussed in section 2.1 the PDS data of H1846-786 could be
contaminated by other Seyfert galaxies present in the field of view,
therefore we first analysed only the LECS and MECS data. In the
0.1-10 keV band a simple canonical power law is the best fit to the
data (see figure 1(d)). This result indicates no need for extra
absorption or the presence of a soft excess. Also the presence of an
iron line is not statistically required by the data; in any case
fixing the line energy at 6.4 keV as expected in the case of the iron
K$_\alpha$ fluorescence line, we obtain an upper limit on the equivalent width
of $\sim$ 200 eV again compatible with the values seen in Seyfert 1s.  

We also tried to put some constrain on the reflection and high energy
cut-off taking into consideration the PDS data and assuming first that
non contaminating sources were present in the field of view. In this
case we estimate R to be 2.6, while  the high energy
cut-off cannot be constrained. 
On the other hand, in a more realistic case, we should consider the contribution of at least
an AGN having a 2-10 keV flux 1/5 that of H1846-786, a photon index of $\Gamma$=1.9 and
a reflection component R=1, to account for contamination in
the PDS data.
In this case, the value of the reflection component measured in H1846-786 is still greater than 2
and the high energy cut-off remains unconstrained.

\section{Conclusion}

BeppoSAX observations of four Seyfert 1 galaxies of the Piccinotti
sample indicate the presence of complex absorption in two objects
(H0557-385 and H0111-149): this absorption is best described by the
combination of two uniform absorbers, one cold and one warm. Only one
object in the sample (H0557-385) requires a soft excess component.
The primary continuum is best described in all sources by a canonical power law with a
high energy cut-off in the range 40 keV to higher than 130 keV.  This
seems to be indicative of a large dispersion in cut-off energies due to
intrinsic differences between sources but also to variations within a
single source; at least in one galaxy (H0557-385) we have some
evidence for a change in cut-off energy. A cold reflection component
is likely present in all sources: here the observed range of
values is large too ranging from less than 0.6 to higher than
2. This may be due to a large dispersion in this parameter over
sources and/or to variations in R in the same source as observed in
the case of H0235-525. In 3 out of 4 objects we find a cold iron line
having equivalent width typical of type 1 objects, i.e. 100-200
eV. The iron line EW is in all cases compatible with the observed
strength of the reflection component.  
In a following paper, we expect to extend this study
to the entire Piccinotti sample of Seyfert 1 galaxies in order to
constrain further the parameter space of the various spectral
components and to study possible correlation between them.

\begin{acknowledgements}
This research has made use of SAXDAS linearized and cleaned event files produced at
the BeppoSAX Science Data Center. We are grateful to P.~O.~Petrucci for useful discussions 
on H0235-525.
\end{acknowledgements}

{}


\begin{thebibliography}{}
\bibitem[]{} Anders, E., Grevesse, N., 1993, Geochim. Cosmochim. Acta 53, 197
\bibitem{} Fairall, A.P., McHardy, I.M., Pye, J.P., 1982, MNRAS, 198, 13
\bibitem{} Fiore, F., Guainazzi, M., Grandi, P., 1998, BeppoSAX Cookbook
\bibitem{} Frontera, F., Costa, E., Dal Fiume, D., Feroci, M., Nicastro, L., Orlandini, M., Palazzi, E., Zavattini, G., 1997, A \& A, 122, 357
\bibitem{} Giommi, P., Bevermann, K., Barr, P., Schwope, A., Tagliaferri, G., Thomas, H.C., 1989, MNRAS, 236, 375
\bibitem{} Guainazzi, M., 2003, A\&A, 903, 910
\bibitem{} Magdziarz, P., Zdziarski, A., 1995, MNRAS, 273, 837
\bibitem{} Malizia, A., Bassani, L., Zhang, K.A., Dean, A.J., Paciesas, W.S., Palumbo, G.G.C., 1999, ApJ, 519, 637
\bibitem{} Perola, G.C., Matt, G., Cappi, M., Fiore, F., Guainazzi, M., Maraschi, L., Petrucci, P.O., Piro, L., 2002, A \& A, 389, 802
\bibitem{} Piccinotti, G.,  Mushotzky, R.F., Boldt, E.A., Holt, S.S., Marshall, F.E., Serlemitsos, P.J., Shafer, R.A., 1982, ApJ, 253, 485
\bibitem{} Remillard, R.A., Bradt, H.V., Buckley, D.A.H., Roberts, W., Schwartz, D.A., Tuohy, I.R., Wood, K., 1986, ApJ, 301,742
\bibitem{} Schartel, N., Schmidt, M., Fink, H.H., Hasinger, G., Tr$\ddot u$mper, J., 1997, A \& A, 320, 696
\bibitem{} Turner, T.J., Netzer, H., Gearge, I.M., 1996, ApJ, 463, 134T
\bibitem{} Turner, T.J., Pounds, K.A., 1989, MNRAS, 240, 833
\bibitem{} Ward, M.J., Elvis, M., Fabiano, G., Carleton, N.P., Willner, S.P., Lawrence, A., 1987, ApJ, 315, 74
\bibitem{} Ward, M.J., Shafer, R.A., 1988, ApJ, 324, 767
\end{thebibliography}
\end{document}